\documentclass{elsart}

\usepackage{graphicx}
\usepackage{amsfonts}

\usepackage{amssymb}

\begin{document}
\begin{frontmatter}
\title{The entanglement character between
atoms in the non-degenerate two photons Tavis-Cummings model}
\author{Guo-feng Zhang\corauthref{cor1}}
\corauth[cor1]{Corresponding author.} \ead{gf1978zhang@163.com}
\author{and Zi-yu Chen}
\address{Department of physics, School
of sciences, \\Beijing University of Aeronautics and Astronautics
(BUAA), \\Beijing 100083, China}

\begin{abstract}
\quad The entanglement character including the so-called sudden
death effect between atoms in the non-degenerate two photons
Tavis-Cummings model is studied by means of concurrence. The results
show that the so-called sudden death effect occurs only for some
kind of initial states. In other words, the phenomenon is sensitive
to the initial conditions. One can expect the resurrection of the
original entanglement to occur in a periodic way following each
sudden death event. The length of the time interval for the zero
entanglement is dependent on not only the degree of entanglement of
the initial state but also the initial state. And the influence of
dipole-dipole interaction and different atomic initial states on
entanglement between atoms are discussed. The sudden death effect
can be weakened by the introducing of dipole-dipole interaction.
\end{abstract}

\end{frontmatter}

\quad Entanglement is one of the most striking features of quantum
mechanics. It plays a fundamental role in almost all efficient
protocols of quantum computation (QC) and quantum information
processing (QIP). In recent years, there has been an ongoing effort
to characterize qualitatively and quantitatively the entanglement
properties and apply then in quantum communication and information.
Many schemes are proposed for generation of two or many particles
entanglement. Among these, the system of atoms interacting with
cavity field is a prominent scheme. Bose \emph{et al}. show that
entanglement can always arise in the interaction of an arbitrarily
large system in any mixed state with a single qubit in a pure
state\cite{bs1,bs2}. They demonstrate this feature using the
Jaynes-Cummings interaction of a two-level atom in a pure state with
a field in a thermal state at an arbitrarily high temperature. Their
method can be also used to study the entanglement between atom and
cavity field in a loss Jaynes-Cummings model (JCM) and between two
identical atoms in Tavis-Cummings model (TCM) \cite{msk}. Tessier
\emph{et al}. made a generalization of the JCM work to TCMS
\cite{tes}. Boukobza \emph{et al}. studied the relation between
entanglement and entropy in JCM \cite{ebo}. Recently, the dynamics
of entanglement in bipartite systems has afresh received great
attention since the work of Yu and Eberly \cite{tyu}, in which the
entanglement between the two particles coupled with two independent
environments became completely vanishing in a finite time. This
surprising phenomenon, contrary to intuition based on experience
about qubit decoherence, intrigues great interests
\cite{kro,zfi,rfl,my,ty1,ty2}. In Ref.\cite{zfi,rfl}, the authors
show that for a special initial state the entanglement disappeared
in a finite time and then revived after a dark period because of the
interaction between the particles, which is different from the works
by Yu and Eberly \cite{tyu} since in Ref.\cite{zfi,rfl} the effects
of interaction between the particles and the couplings to the same
environment have been discussed extensively. In Ref.\cite{msk}, the
dipole-dipole interaction is ignored and only the case that when two
atoms are initially in separate state is studied. What will happen
if the atoms are initially in a entangled state and the
dipole-dipole interaction is turned on. Thus, the purpose of this
paper is to examine the entanglement character between atoms in the
non-degenerate two photons TCM (TPTCM) and investigate the sudden
death effect when the dipole-dipole interaction is turned off. The
entanglement evolvement for a general case when considering the
dipole-dipole interaction is given.

\quad Consider the TPTCM where two identical two-level atoms are in
a two-mode cavity field. By tracing cavity modes we are forcibly
creating a two-qubit scenario, and various measures of entanglement
are available. For a pair of qubits, all of them are equivalent, in
the sense that when any one of them indicates no entanglement
(separable states), the others also indicate no entanglement.
Throughout the paper we will use Wootters's \cite{wkw} concurrence
$C(\rho)$ as the conveniently normalized entanglement measure
$(1\geq C\geq0)$.

\quad The dipole-dipole interaction of the atoms can not be
neglected when the distance of the atoms is less than the wave
length in the cavity. The Hamiltonian can be given
\begin{eqnarray}
H=\sum_{i=a}^{b}\omega_{i}a_{i}^{\dagger}a_{i}&+&\frac{1}{2}\omega_{0}\sum_{l=A}^{B}\sigma_{l}^{z}+g\sum_{l=A}^{B}(a_{a}^{\dagger}a_{b}^{\dagger}\sigma_{l}^{-}+a_{a}a_{b}\sigma_{l}^{+})\nonumber\\&+&\Omega
(\sigma_{A}^{+}\sigma_{B}^{-}+\sigma_{B}^{+}\sigma_{A}^{-}),
\end{eqnarray}
where $a_{i}$ and $a_{i}^{\dagger}$ $(i=a,b)$ denote the
annihilation and creation operators of the frequency $\omega_{i}$
quantization field. $\omega_{0}$ is the atomic Rabi frequency.
$\sigma^{l}_{z}=|e\rangle_{ll}\langle e|-|g\rangle_{ll}\langle g|$,
$\sigma^{l}_{+}=|e\rangle_{ll}\langle g|$,
$\sigma^{l}_{-}=|g\rangle_{ll}\langle e|$ are the atomic operators
with $|e\rangle_{l}$ and $|g\rangle_{l}$ being the excited and
ground states of the $l$th atom $(l=A,B)$. $g$ is the coupling
constant between atom and field, $\Omega$ is dipole-dipole coupling
strength between atoms.

\quad For greatest simplicity, we assume that both cavities modes
are prepared initially in the vacuum state
$|0_{a}\rangle\otimes|0_{b}\rangle=|00\rangle$ and the two atoms are
in a pure entangled state specified below. This allows a uniform
measure of quantum entanglement---concurrence. In principle, there
are six different concurrences that provide information about the
overall entanglements that may arise. With an obvious notation we
can denote these as $C^{Aa}$, $C^{Ab}$, $C^{Ba}$, $C^{Bb}$,
$C^{AB}$, $C^{ab}$. Here we confine our attention to $C^{AB}$.

\quad We assume that the two atoms are initially in a partially
entangled atomic pure state that is a combination of Bell states
usually denoted $|\Psi^{\pm}\rangle$, we have
\begin{equation}
|\Psi_{atom}(0)\rangle=\cos[\alpha]|eg\rangle+\sin[\alpha]|ge\rangle,
\end{equation}
with the first index denoting the state of atom $A$ and the second
denoting the state of atom $B$. Thus the initial state for the
system (1) can be given by
\begin{equation}
|\Psi(0)\rangle=|\Psi_{atom}(0)\rangle\otimes|00\rangle=\cos[\alpha]|eg00\rangle+\sin[\alpha]|ge00\rangle.
\end{equation}
The solution of the system in terms of the standard basis can be
written as
\begin{equation}
|\Psi(t)\rangle=x_{1}|eg00\rangle+x_{2}|ge00\rangle+x_{3}|gg11\rangle,
\end{equation}
where $\omega_{0}=\omega_{a}+\omega_{b}$ is assumed. The
corresponding coefficients is given by
\begin{eqnarray}
x_{1}&=&\frac{\Lambda}{4}[\theta_{+}(L_{+}e^{i\kappa T}-L_{-})+2\Xi\theta_{-}],\nonumber\\
x_{2}&=&\frac{\Lambda}{4}[\theta_{+}(L_{+}e^{i\kappa T}-L_{-})-2\Xi\theta_{-}],\nonumber\\
x_{3}&=&\frac{\Lambda\theta_{+}}{\kappa}(1-e^{i\kappa T}).
\end{eqnarray}
with $\theta_{\pm}=\cos[\alpha]\pm\sin[\alpha]$ and $T=gt$.
$\Lambda=e^{-i\kappa L_{+}T/2}$, $\Xi=e^{i(3L_{+}-2)\kappa T/2}$,
$L_{\pm}=\frac{\epsilon}{\kappa}\pm1$, $\epsilon=\frac{\Omega}{g}$
and $\kappa=\sqrt{8+\epsilon^{2}}$. The information about the
entanglement of two atoms is contained in the reduced density matrix
$\rho^{AB}$ which can be obtained from (4) by tracing out the
photonic part of the total pure state. In the standard basis
$\{|1,1\rangle,|1,0\rangle,|0,1\rangle,|0,0\rangle\}$, the density
matrix $\rho^{AB}$ can be expressed as
\begin{equation}
\rho^{AB}=\left(%
\begin{array}{cccc}
  0 & 0 & 0 & 0 \\
  0 & |x_{1}|^{2} & x_{1}x_{2}^{*} & 0 \\
  0 & x_{2}x_{1}^{*} & |x_{2}|^{2} & 0 \\
  0 & 0 & 0 & |x_{3}|^{2} \\
\end{array}%
\right),
\end{equation}
it can be shown that the concurrence associated with the density
matrix is given by
\begin{equation}
C^{AB}(t)=2\max\{|x_{1}x_{2}^{*}|,0\}.
\end{equation}
\begin{figure}[h]
\begin{center}
\includegraphics[width=8 cm]{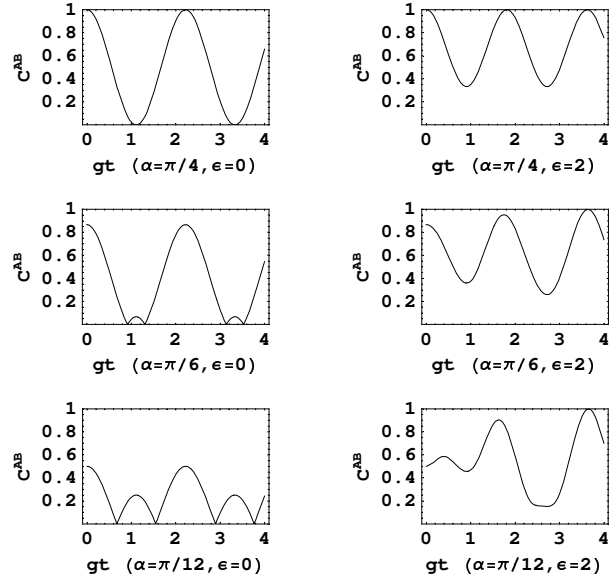}
\caption{The concurrence for atom-atom entanglement with the initial
atomic state
$|\Psi_{atom}(0)\rangle=\cos[\alpha]|eg\rangle+\sin[\alpha]|ge\rangle$.}
\end{center}
\end{figure}
Fig.1 shows the entanglement evolvement for the different $\alpha$
when $\epsilon=0$ and $\epsilon=2$. It is seen that entanglement can
be strengthened by introducing the dipole-dipole interaction.
Although Eq. (4) is not always an entangled state for atoms, with
the increasing $\epsilon$ the coefficient of unpolarized state
$x_{3}$ will be smaller until to be zero, thus the system state
$|\Psi(t)\rangle\rightarrow x_{1}|eg00\rangle+x_{2}|ge00\rangle$
which is entangled for the two atoms. We can also see that the
evolvement period gets shorter with the increasing $\epsilon$. This
is very easily understood since the period is governed by
$\cos[\sqrt{8+\epsilon^{2}}gt]$. From the figures, we can know that
dipole-dipole interaction can cause the entanglement be maximum
although the initial entanglement may be very weak. The initial
entanglement is determined by Eq.(2),
$C(0)=2\cos[\alpha]\sin[\alpha]=\sin[2\alpha]$ which is a fixed
value. $\epsilon=0$,
$C(t)=\frac{1}{4}(-1+3\sin[2\alpha]+\cos[\sqrt{8}gt](1+\sin[2\alpha]))$,
$C(t)_{max}=C(0)$. But for $\epsilon\neq0$, $C(t)$ is dependent on
$\epsilon$, and it can arrive at the maximum value 1 for any
$\epsilon$ and $\alpha$ at some evolvement time points which depend
on $\epsilon$ and are independent of $\alpha$. These can be seen
from the right panel of Fig.1. In Eq.(5), $x_{1}$ , $x_{2}$ can
arrive $\sqrt{2}/2$ and $x_{2}$ can be zero at the same time, for
that case the entanglement is 1. This can be also understood from
the physics since the independence of atoms will be weakened by the
dipole-dipole interaction, thus the initial entangled state will be
more entangled.

\quad Now we assume that the initial state for the total system is
in a combination of the other two Bell states $|\Phi^{\pm}\rangle$
\begin{equation}
|\Phi(0)\rangle=\cos[\alpha]|ee00\rangle+\sin[\alpha]|gg00\rangle,
\end{equation}
in which case the state of the total system at time $t$ can be
expressed in the standard basis
\begin{equation}
|\Phi(t)\rangle=x_{1}|ee00\rangle+x_{2}|gg00\rangle+x_{3}|ge11\rangle+x_{4}|eg11\rangle+x_{5}|gg22\rangle,
\end{equation}
where the coefficients are now given by
\begin{eqnarray}
x_{1}&=&\frac{\Gamma}{4}(\frac{\epsilon}{\eta}M_{-}+M_{+}+2e^{i(\epsilon+\eta)T/2}),\nonumber\\
x_{2}&=&e^{i\lambda T}\sin[\alpha],\nonumber\\
x_{3}&=&x_{4}=\frac{\Gamma M_{-}}{\eta},\nonumber\\
x_{5}&=&\frac{\Gamma}{4}(\frac{\epsilon}{\eta}M_{-}+M_{+}-2e^{i(\epsilon+\eta)T/2}).
\end{eqnarray}
with $\lambda=\frac{\omega}{g}$, $\eta=\sqrt{16+\epsilon^{2}}$,
$\Gamma=\cos[\alpha]e^{-i(2\lambda+\epsilon+\eta)T/2}$ and
$M_{\pm}=1\pm e^{i\eta T}$. In the basis of $|ee\rangle$,
$|eg\rangle$, $|ge\rangle$ and $|gg\rangle$ the reduced density
matrix $\rho^{AB}$ is now found to be
\begin{equation}
\rho^{AB}=\left(%
\begin{array}{cccc}
  |x_{1}|^{2} & 0 & 0 & x_{1}x_{2}^{*} \\
  0 & |x_{4}|^{2} & x_{4}x_{3}^{*} & 0 \\
  0 & x_{3}x_{4}^{*} & |x_{3}|^{2} & 0 \\
  x_{2}x_{1}^{*} & 0 & 0 & |x_{2}|^{2}+ |x_{5}|^{2}\\
\end{array}%
\right).
\end{equation}

\begin{figure}[h]
\begin{center}
\includegraphics[width=8 cm]{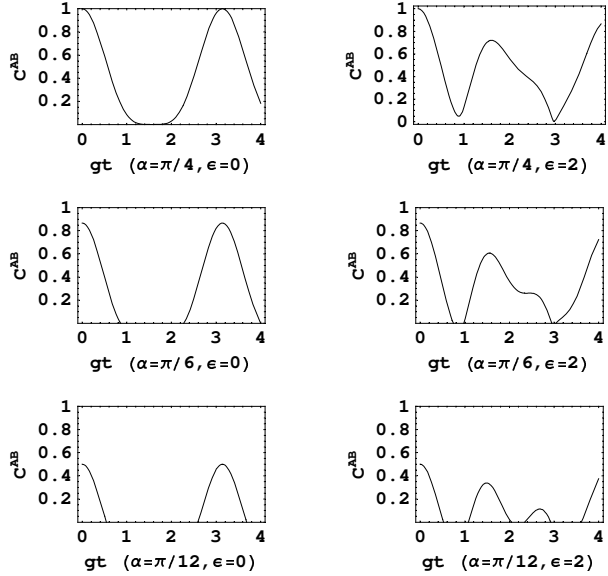}
\caption{The concurrence for atom-atom entanglement with the initial
atomic state
$|\Psi_{atom}(0)\rangle=\cos[\alpha]|ee\rangle+\sin[\alpha]|gg\rangle$.}
\end{center}
\end{figure}

And the corresponding concurrence for this matrix can be obtained
based on Ref.\cite{wkw}. The numerical results are plotted in Fig.2.
Unlike the previous case, figure 2 show that entanglement can fall
abruptly to zero (the two lower curves in the figure), and it will
remain to be zero for a period of time before entanglement recovers.
The length of the time interval for the zero entanglement is
dependent on the degree of entanglement of the initial state. The
smaller the initial degree of entanglement is, the longer the state
will stay in the disentangled separable state. Compared the right
panel with the left one, one will find that the dipole-dipole
interaction can cause the length of the time interval for the zero
entanglement be shorter. When $\epsilon=0$, let us first review the
Hamiltonian Eq. (1), obviously the dynamics of the two independent
two-level atoms is determined by the interaction terms in Eq. (1),
which is the charge of the energy transfer between the system and
field. We think it may be the energy transfer that leads to the
entanglement sudden death (ESD). Originally ESD comes from the
cutoff in the definition of concurrence. In our own points the main
obstacle of explaining these phenomena is that it is unclear what
the physical meanings of the concurrence are. Recently a great deal
of works are denoted to the understanding from the energy of the
system \cite{gto}. More recently a physical interpretation of
concurrence for the bipartite systems has been provided based on the
Casimir operator in \cite{akl}. It maybe open another door to
understand concurrence as a physical quantity.

\quad In summary, we have investigated the entanglement character
existed in a non-degenerate two photons Tavis-Cummings model. We
found that for the two different initial state of atoms, the
entanglement evolvement appear dramatic difference. The length of
the time interval for the zero entanglement is dependent on not only
the degree of entanglement of the initial state but also the initial
state. And the introducing of dipole-dipole interaction can cause
the entanglement to be higher and weaken the difference of
entanglement evolvement between two kind different initial state of
atoms.

\textbf{Acknowledgements}

\quad This work was supported by the National Science Foundation
of China under Grants No. 10604053.

\end{document}